\documentclass[12pt]{article}

\usepackage{amsmath, amssymb}
\usepackage{amsthm}
\usepackage{graphicx}
\usepackage{tikz}

\usepackage[
  left=1.15in,
  right=1.15in,
  top=1.4in,
  bottom=1.4in
]{geometry}

\usepackage{pgfplots}
\pgfplotsset{compat=1.18}

\newtheorem{theorem}{Theorem}
\newtheorem{lemma}[theorem]{Lemma}
\newtheorem{proposition}[theorem]{Proposition}

\theoremstyle{remark}

\numberwithin{theorem}{section} 

\theoremstyle{definition}

\theoremstyle{remark}
\newtheorem*{remark}{Remark}

\newcommand{\Rpos}{\mathbb{R}_{>0}}

\begin{document}

\title{Alignment in products of positive matrices and Birkhoff's contraction formula: an elementary direct derivation}

\author{Evgenij Kritchevski}

\maketitle
\begin{abstract} We revisit the classical problem of alignment in products of positive matrices. A direct analysis yields a sharp product misalignment bound, providing an elementary route to alignment and Birkhoff's contraction formula.
\end{abstract}

\section{Introduction}
We begin by recalling Birkhoff's contraction formula in the finite-dimensional setting and its consequences for alignment of columns and rows in products of positive matrices. In \cite{Birkhoff1957}, Birkhoff introduced a geometric approach to Perron--Frobenius theory  by showing that positive operators act as contractions in the Hilbert projective metric
$$d_n(x,y):=\log \max_{i,j}\frac{y_i x_j}{y_j x_i}\qquad x,y\in\Rpos^{n}.$$
Remarkably, the optimal contraction coefficient is given by an explicit closed-form expression in terms of the matrix entries. For $A\in\Rpos^{m\times n}$, Birkhoff's contraction formula states that
$$\sup_{\substack{x,y\in \Rpos^n \\ d_n(x,y)\neq 0}}\frac{d_m(Ax,Ay)}{d_n(x,y)}=\tau(A),$$
where
$$\tau(A):=\frac{\sqrt{R(A)}-1}{\sqrt{R(A)}+1}\qquad R(A):=\max_{i,j,k,\ell}
   \frac{a_{ik}\, a_{j\ell}}{a_{i\ell}\, a_{jk}}.$$
Both $R(A)$ and $\tau(A)$ measure column (and row) misalignment. When the rank of $A$ is $1$, alignment is perfect and we have $R(A)=1$ and $\tau(A)=0$. Otherwise, $R(A)>1$ and $0<\tau(A)<1$. 

Alignment in products is then the effect of repeated contractions. Since contraction coefficients are submultiplicative, we have
\[
\tau(A_N\cdots A_2A_1)\leq \prod_{r=1}^{N}\tau(A_r)
\]
for every well-defined product of positive matrices. Consequently, in the presence of a uniform upper bound on \(R(A_r)\), or more generally whenever
\[
\lim_{N\to\infty}\prod_{r=1}^{N}\tau(A_r)= 0,
\]
one concludes that
\[
\lim_{N\to\infty}\tau(A_N\cdots A_2A_1)=0.
\]
Therefore the columns (and likewise the rows) of the product \(A_N\cdots A_2A_1\) become asymptotically aligned.

In this note, we take a conceptually different route to these classical matrix results. Instead of relying on contraction arguments in the Hilbert projective metric, we begin with the problem of controlling alignment in products of positive matrices. This naturally leads to the sharp product misalignment bound
$$R(AB)\leq\left(\frac{1+\sqrt{R(A)R(B)}}{\sqrt{R(A)}+\sqrt{R(B)}}\right)^{\!2} $$
from which both the submultiplicative inequality
$$
\tau(AB)\le \tau(A)\tau(B)
$$
and Birkhoff's contraction formula are derived. In fact, the product misalignment bound, the submultiplicativity of $\tau$, and Birkhoff's contraction formula are all equivalent. Among these equivalent statements, the product misalignment bound appears to admit the most direct elementary proof.

Since Birkhoff's seminal work, the contraction theory has been extensively developed and generalized to abstract cone settings. For a comprehensive treatment, see \cite{LemmensNussbaum2012} and the references therein. In the finite dimensional setting, several elementary derivations of Birkhoff's contraction formula are available; see, for example, \cite{CavazosCadena2003,Carroll2004}. Our approach gives an elementary self contained explanation of alignment in products of positive matrices by deriving the submultiplicative inequality for $\tau$ directly, without appealing to its interpretation as a contraction coefficient. In addition, we obtain an alternative elementary proof of Birkhoff's formula. 
\section{The product misalignment bound}
We seek a sharp bound for $R(AB)$ in terms of $R(A)$ and $R(B)$. We start by analysing the case where $A$ and $B$ are $2$ by $2$ matrices. We note that $R(AB)$ is invariant under the following operations: swapping or rescaling the rows of $A$, swapping or rescaling the columns $B$. Also note for any $M\in\Rpos^{2\times 2}$ with $\det(M)\geq 0$, we have $\displaystyle R(M)=\frac{m_{11}m_{22}}{m_{12}m_{21}}$. Therefore, we may assume without loss of generality that $\det(A)\geq0$, $\det(B)\geq 0$, 
\[
A=\begin{pmatrix}1 & u \\[2mm] 1 & \alpha u \end{pmatrix},
\qquad
B=\begin{pmatrix}1 & 1 \\[2mm] v & \beta v \end{pmatrix},
\]
where
\[
\alpha=R(A)\ge1,\qquad \beta=R(B)\ge1.
\]
Then
\[
AB=
\begin{pmatrix}
1+t & 1+\beta t \\[2mm]
1+\alpha t & 1+\alpha\beta t
\end{pmatrix},
\qquad t=uv>0.
\]

Since $\det(AB)\geq 0$, 
\[
R(AB)=f_{\alpha,\beta}(t),
\qquad
f_{\alpha,\beta}(t)
:=\frac{(1+t)(1+\alpha\beta t)}{(1+\beta t)(1+\alpha t)},
\quad t>0.
\]
The task of bounding $R(AB)$ therefore reduces to maximizing the one-variable function
$f_{\alpha,\beta}(t)$. Differentiating its logarithm yields
\[
\frac{d}{dt}\log f_{\alpha,\beta}(t)
=\frac{(\alpha-1)(\beta-1)\,(1-\alpha\beta t^2)}
{(1+t)(1+\alpha t)(1+\beta t)(1+\alpha\beta t)}.
\]
Thus $f_{\alpha,\beta}$ is maximized at 
$$t^*:=\frac{1}{\sqrt{\alpha\beta}}.$$ 
A direct substitution and simplification yield the maximal value
\[
f_{\alpha,\beta}(t^*)=\Phi(\alpha,\beta),
\]
where
\[
\Phi(\alpha,\beta)
:=\left(\frac{1+\sqrt{\alpha\beta}}{\sqrt{\alpha}+\sqrt{\beta}}\right)^2.
\]
Thus we have established:
\begin{proposition}[Product misalignment bound in dimension $2$]\label{R2by2}
 For $A,B\in\Rpos^{2\times 2}$, 
$$R(AB)\leq \Phi(R(A),R(B)).$$
\end{proposition}

We next prove that the product misalignment bound holds in arbitrary dimension. For $x,y\in\Rpos^n$, define
\[
s_n^+(x,y):=\max_i \frac{y_i}{x_i} \qquad 
s_n^-(x,y):=\min_i \frac{y_i}{x_i} \qquad \delta_n(x,y):=\frac{s^+(x,y)}{s^-(x,y)}.
\]
Then
$$\delta_n(x,y)= \max_{i,j}\frac{y_i x_j}{y_j x_i}.$$
Define $F_n : (\mathbb{R}^n_{>0})^4 \to \mathbb{R}_{>0}$ by
\[
F_n(u,v;x,y):=\frac{(u\cdot x)(v\cdot y)}{(u\cdot y)(v\cdot x)},
\]
where $\cdot$ denotes the standard Euclidean dot product.
\begin{lemma}\label{reduction2d} For any $u,v,x,y\in\Rpos^n$, there exist $u',v',x',y'\in\Rpos^2$ such that
$$\delta_2(u',v')\leq \delta_n(u,v)\qquad \delta_2(x',y')\leq \delta_n(x,y)\qquad  F_n(u,v;x,y)\leq F_2(u',v';x',y').$$ 
\end{lemma}

\begin{proof}
Denote \(a^{\pm}=s_n^\pm(u,v)\) and \(b^{\pm}=s_n^\pm(x,y)\). If $a^-=a^+=a$, the Lemma holds trivially with $u'=(1,1)$, $v'=(a,a)$, $x'=(1,1)$ and $y'=(b^-,b^+)$. So we assume from now on that $a^-<a^+$. Note that, with all other variables fixed, \(F_n(u,w;x,y)\) is a linear-fractional function of each component \(w_j\), hence monotone.  If
\[
a^-<\frac{v_j}{u_j}<a^+
\]
for some index $j$, we replace \(v_j\) by one of \(a^-u_j\) and \(a^+u_j\) so that
\(F_n(u,v;x,y)\) does not decrease; such a choice is possible by
monotonicity. Repeating this coordinate update successively for every remaining index $j$ such that $\displaystyle a^-<\frac{v_j}{u_j}<a^+$, while keeping all previously modified coordinates fixed, produces \(v^*\in\Rpos^n\) such that
\[
F_n(u,v;x,y)\le F_n(u,v^*;x,y),\qquad
s_n^\pm(u,v^*)=a^\pm,
\]
and
\[
\frac{v_i^*}{u_i}\in\{a^-,a^+\}
\qquad\text{for all }i.
\]
Define the index sets
\[
I_\pm:=\left\{\,i:\frac{v_i^*}{u_i}=a^\pm\right\},
\]
positive scalars
\[
u^\pm:=\sum_{i\in I_\pm}u_i,\qquad
v^\pm:=a^\pm u^\pm,\qquad
x^\pm:=\frac1{u^\pm}\sum_{i\in I_\pm}u_i x_i,\qquad
y^\pm:=\frac1{u^\pm}\sum_{i\in I_\pm}u_i y_i,
\]
and vectors in $\Rpos^2$
\[
u':=(u^-,u^+),\quad
v':=(v^-,v^+),\quad
x':=(x^-,x^+),\quad
y':=(y^-,y^+).
\]
Then $\delta_2(u',v')=\delta_n(u,v)$ and $F_2(u',v';x',y')=F_n(u,v^*;x,y)$ by construction.
%Then $u'\cdot x'=u\cdot x$, $v'\cdot y'=v^*\cdot y$, $u'\cdot y'=u\cdot y$, $v'\cdot x'=v^*\cdot x$ and therefore $F_2(u',v';x',y')=F_n(u,v^*;x,y)$.
Moreover, writing 
$$\frac{y^\pm}{x^\pm}=\frac{1}{x^\pm u^\pm}\sum_{i\in I_\pm}x_iu_i \frac{y_i}{x_i},$$
we see that $\displaystyle\frac{y^\pm}{x^\pm}$ is a weighted average of $\displaystyle \frac{y_i}{x_i}$. Since $\displaystyle b^- \leq \frac{y_i}{x_i} \leq b^+$, we must have
$\displaystyle b^- \leq \frac{y^\pm}{x^\pm} \leq b^+$ as well. It follows that $\displaystyle\delta_2(x',y')\leq \frac{b^+}{b^-}=\delta_n(x,y)$, completing the proof.

\end{proof}

\begin{remark} From the proof, we actually get $\delta_2(u',v')=\delta_n(u,v)$. It is possible to modify the construction of $x'$ and $y'$ to get $\delta_2(x',y')= \delta_n(x,y)$ as well, but the proof becomes slightly longer. For our purpose, the weaker conclusions $\delta_2\leq \delta_n$ are sufficient.
\end{remark}

\begin{theorem}[Product misalignment bound in arbitrary dimension]\label{Rany}
For $A\in\Rpos^{m\times n}$ and  $B\in\Rpos^{n\times p}$,
$$R(AB)\leq \Phi(R(A),R(B)).$$
\end{theorem}
\begin{proof}
Let $C=AB$. Then
$$\frac{c_{ik}\, c_{j\ell}}{c_{i\ell}\, c_{jk}}=F_n(u,v;x,y)$$
where
\[
u=\text{row }i\text{ of }A,\quad v=\text{row }j\text{ of }A,\qquad
x=\text{column }k\text{ of }B,\quad y=\text{column }\ell\text{ of }B.
\]
By Lemma \ref{reduction2d}, there exist $u',v',x',y'\in\Rpos^2$ such that
$$\delta_2(u',v')\leq \delta_n(u,v)\qquad \delta_2(x',y')\leq \delta_n(x,y)\qquad F_n(u,v;x,y)\leq F_2(u',v';x',y').$$
Let $A'$ be the $2$ by $2$ matrix with rows $u'$ and $v'$ and let  $B'$ be the $2$ by $2$ matrix with columns $x'$ and $y'$. Let $C'=A'B'$. Then
$$F_2(u',v';x',y')=\frac{c'_{11}\, c'_{22}}{c'_{12}\, c'_{21}} \leq R(C').$$
Proposition \ref{R2by2} gives
$$R(C')\leq\Phi(R(A'),R(B')).$$
Since $\Phi$ is increasing in each variable and 
$$R(A')=\delta_2(u',v')\leq \delta_n(u,v)\leq R(A) \qquad R(B')=\delta_2(x',y')\leq \delta_n(x,y)\leq R(B),$$
we obtain
$$\Phi(R(A'),R(B'))\leq \Phi(R(A),R(B)).$$
Combining the chain of inequalities yields
$$\frac{c_{ik}\, c_{j\ell}}{c_{i\ell}\, c_{jk}} \leq \Phi(R(A),R(B))$$
and taking the $\max$ over all $i,j,k,\ell$ completes the proof.
\end{proof}

\section{Submultiplicativity of $\tau$ and Birkhoff's formula}
\begin{theorem}[Submultiplicativity of $\tau$]\label{submult}
For $A\in\Rpos^{m\times n}$ and  $B\in\Rpos^{n\times p}$,
$$\tau(AB)\leq \tau(A)\tau(B).$$
\end{theorem}
\begin{proof}
With $S(A):=\sqrt{R(A)}$, the product misalignment bound gives
$$S(AB)\leq \Psi(S(A),S(B))\qquad \Psi(a,b):=\frac{1+ab}{a+b}.$$
Let $\displaystyle g(s):=\frac{s-1}{s+1}$, $s\geq 1$. The identity
$$g(\Psi(a,b))=\frac{\frac{1+ab}{a+b} -1}{\frac{1+ab}{a+b} +1}=\frac{1+ab-a-b}{1+ab+a+b}=\frac{(a-1)(b-1)}{(a+1)(b+1)}=g(a)g(b)$$
and the fact that $g$ is increasing on $[1,\infty)$ yield
$$\tau(AB)=g(S(AB))\leq g(\Psi(S(A),S(B)))=g(S(A))g(S(B))=\tau(A)\tau(B).$$
\end{proof}
\begin{remark} Since $\Psi$ is a Möbius transformation in each variable with fixed points $\pm 1$, the change of variables
$\displaystyle\tau=\frac{S-1}{S+1}$ and hence the definition of $\tau(A)$ appear naturally. The change of variables also shows that submultiplicativity of $\tau$ implies the product distortion bound.
\end{remark}

\begin{theorem}[Birkhoff's contraction formula]
For $A\in\Rpos^{m\times n}$,
$$\sup_{\substack{x,y\in \Rpos^n \\ d_n(x,y)\neq 0}}\frac{d_m(Ax,Ay)}{d_n(x,y)}=\tau(A),$$
\end{theorem}
\begin{proof}
Consider the matrix $B\in\Rpos^{n\times 2}$ whose columns are $x$ and $y$. Then
$$d_n(x,y)=\log R(B),\qquad d_m(Ax,Ay)=\log R(AB).$$
In this logarithmic setting, the product distortion bound takes the form
\[
d_m(Ax,Ay)\;\le\;\Theta\!\left(d_n(x,y)\right),
\]
where
\[
\Theta(h)
   :=2\log\!\left(\frac{1+p\,e^{h/2}}{\,p+e^{h/2}}\right),
   \qquad p:=\sqrt{R(A)}.
\]
A direct computation gives
\[
\Theta'(h)=\frac{(p^{2}-1)e^{h/2}}{(1+pe^{h/2})(p+e^{h/2})},
\qquad
\Theta''(h)
=-\frac{(p^{2}-1)e^{h/2}p(e^{h}-1)}
{2(1+pe^{h/2})^{2}(p+e^{h/2})^{2}}.
\]
It follows that $\Theta$ is increasing and concave on $[0,\infty)$, and satisfies
\[
\Theta(0)=0, \qquad 
\Theta'(0)=\frac{p-1}{p+1}=\tau(A).
\]
By concavity, $\Theta$ lies below its tangent line at the origin,
$
\Theta(h)\;\le\;\tau(A)\,h,
$
which yields Birkhoff's contraction
\[
d_m(Ax,Ay)
   \;\le\;\tau(A)\, d_n(x,y).
\]

It remains to prove sharpness of $\tau(A)$. Pick indices $1\leq i,j\leq m$ and $1\leq k,\ell\leq n$ for which
$$\displaystyle R(A)=\frac{a_{ik}\, a_{j\ell}}{a_{i\ell}\, a_{jk}}$$
and let
$$A':=\begin{pmatrix}a_{ik} & a_{i\ell} \\[2mm] a_{jk} & a_{j\ell}\end{pmatrix}.$$
Then $R(A')=R(A)$. The optimization argument in the proof of Proposition \ref{R2by2} guarantees that, for any $\beta > 1$, there exists $B'\in\Rpos^{2\times 2}$ such that $R(B')=\beta$ and $R(A'B')=\Phi(R(A),\beta)$. For $\epsilon>0$, construct the matrix $B_\epsilon\in\Rpos^{n\times 2}$ whose rows $k$ and $\ell$ are given respectively by $(b'_{11},b'_{12})$ and $(b'_{21},b'_{22})$, and whose remaining rows are all $(\epsilon b'_{11}, \epsilon b'_{12})$. Then $R(B_\epsilon)=\beta$ and $\lim_{\epsilon\to 0}R(AB_\epsilon)=\Phi(R(A),\beta)$. Translated to the Hilbert metric setting, this means that, for any $h>0$, we can construct 
$x_\epsilon,y_\epsilon\in\Rpos^n$ such that $d_n(x_\epsilon,y_\epsilon)=h$ and $\lim_{\epsilon\to 0} d_m(Ax_\epsilon,Ay_\epsilon)=\Theta(h)$. Now suppose that $0<t<\tau(A)$. By concavity of $\Theta$, there exists $h>0$ such that $\Theta(h)>th$. Then $d_m(Ax_\epsilon,Ay_\epsilon)>th$ for sufficiently small $\epsilon>0$. Therefore $\tau(A)$ cannot be replaced by a smaller contraction constant, completing the proof.
\end{proof}

Since Birkhoff's contraction formula implies the submultiplicativity of $\tau$, while the latter is equivalent to the product misalignment bound via a change of variables, the three statements are equivalent.

%Bibliography

\end{document}